\documentstyle[prl,aps,amsfonts]{revtex}
\input{psfig.sty}
\begin{document}
\draft
\title{Meson Structure in a Relativistic Many-Body Approach}

\author{Felipe J. Llanes-Estrada and Stephen R. Cotanch}
\address{Department of Physics, North Carolina State
University, Raleigh, North Carolina 27695-8202}
\date{\today}
\maketitle

\begin{abstract}

Results from an extensive relativistic many-body analysis utilizing a
realistic effective QCD Hamiltonian are presented for the meson spectrum.
A comparative numerical study of the BCS, TDA and RPA treatments
provides new, significant insight into the condensate structure of the
vacuum, the chiral symmetry governance of the pion and the meson
spin, orbital and flavor mass splitting contributions.  In
contrast to a previous glueball application, substantial quantitative
differences are computed between TDA and RPA for the
light quark sector with the pion emerging as a Goldstone
boson only in the RPA.
\end{abstract}

\pacs{12.39.Pn, 12.39Mk, 11.10.St, 12.39Ki,12.40.Yx}

Common to the diverse areas of condensed matter, molecular, atomic and
nuclear physics is the routine implementation of many-body techniques
such as the Bardeen, Cooper, Schrieffer (BCS), Tamm-Dancoff (TDA) and Random
Phase Approximation (RPA) methods.  Particle physics, with an inherent
few-body nature, has generally been devoid of such applications even though
hadronic structure, requiring a
relativistic QCD description, is an extremely
challenging  many-body problem.  The purpose of the present letter is to
report a  comparative study documenting
the  powerful utility of the above techniques for hadronic systems and to
detail new, important meson structure results clarifying the nature of
spin splittings and role of chiral symmetry.  The equations of motion, while
numerically solvable, exhibit a richness and complexity beyond the simple
two-body equations such as the Bethe-Salpeter or generalized Schr\"{o}dinger
schemes.  We find that both TDA and RPA
solutions to an approximate QCD Hamiltonian with linear confinement
reproduce the meson spectrum except for the pion,
where only the RPA reasonably describes the mass and decay constant due to
proper implementation of chiral symmetry.

This work complements our previous many-body treatment \cite{NCSU1} of the
gluonic sector in which the lattice gauge ``measurements'' were reproduced.
Our collaborative program seeks to develop a rigorous effective
Hamiltonian from QCD and then to comprehensively investigate hadronic
structure by systematic, accurate diagonalization utilizing controllable
approximations.  Ref. \cite{NCSU3} details our renormalization program,
based upon a continuous cut-off regularization and similarity
transformation.  That work addressed only the gluon sector but similar
effort is currently in progress for the quark sector.  Accordingly, this
paper presents many-body solutions for only the
unrenormalized effective Hamiltonian.  The starting
point is the approximate QCD quark Hamiltonian in the Coulomb gauge

\begin{equation}  H = \int d \vec{x} \Psi ^{\dagger} _q (\vec{x})
(-i\vec{\alpha}\cdot \vec{\nabla} + \beta m_q )
\Psi_q(\vec{x}) - \frac{1}{2} \int d \vec{x} d \vec{y}
\rho^a(\vec{x})V(\arrowvert \vec{x} -
\vec{y} \arrowvert) \rho^a(\vec{y}),
\end{equation}

\noindent
involving the quark field $\Psi_q(\vec{x})$, current quark mass $m_q$ and
color density
$\rho^a(\vec{x}) =
\Psi_q^{\dagger}(\vec{x})T^a\Psi_q(\vec{x})$.
Coupling to the gluonic sector is omitted and
the Faddeev-Popov
determinant is replaced by its lowest order unit value.
Consistent with our previous
work \cite{NCSU1}, the confining potential is a linear
interaction,
$V =
\sigma
\arrowvert
\vec{x} - \vec{y} \arrowvert $,
rather than the harmonic oscillator
\cite{Orsay,Lisbon} since lattice gauge theory generates this form with
slope (string tension)  $\sigma =0.18$  $GeV^2$.  Instead of a simplified
gap differential equation for the  harmonic oscillator potential, we  solve
a  numerically quite sensitive nonlinear integral equation (see below) and
reproduce earlier results \cite{Adler}.  The density-density two-body form
permits only color singlets in the physical spectrum as
other $SU_{c}(3)$ representations are shifted to infinite energy.

Next we introduce our first many-body improvement by performing a
BCS rotation (similarity transformation) from the bare (undressed) quark
basis to an improved quasi-particle basis.  This entails rotated
spinors in the quark

\begin{equation} U_{\lambda}(\vec{k}) = \frac{1}{\sqrt{2}} \left[
\begin{array}{c}
\sqrt{1+\sin\phi(k)}\:\: \chi_{\lambda}    \\ \sqrt{1-\sin \phi(k)}\:\:
\vec{\sigma} \cdot \hat{k}\:\: \chi_{\lambda} \end{array} \right],
\quad
V_{\lambda}(\vec{k}) = \frac{1}{\sqrt{2}}\left[ \begin{array}{c}
-\sqrt{1-\sin \phi(k)} \:\:\vec{\sigma}\cdot \hat{k} \:\:\chi_{\lambda} \\
\sqrt{1+\sin \phi(k)} \:\:\chi_{\lambda}\end{array} \right], \end{equation}
field expansion with
quasi-particle operators B, D instead of bare operators b, d
(see ref. \cite{Lisbon} for details)

\begin{equation}
\Psi_q=\sum_{\lambda} \int \frac{d\vec{k}}{(2\pi)^3} \left[
U_{\lambda}(\vec{k})B_{\lambda}(\vec{k}) +
V_{\lambda}(-\vec{k})D^{\dagger}_{\lambda}(-\vec{k})
\right]  e^{i\vec{k} \cdot \vec{x}}.
\end{equation}
The spin state is denoted by $\lambda$ and color indices are suppressed.
The gap angle
$\phi(k)$ governs the BCS vacuum,
$\arrowvert
\Omega
\rangle$, defined by $B_\lambda
\arrowvert
\Omega
\rangle = D_\lambda
\arrowvert \Omega \rangle =0 $.  This vacuum, a coherent state
containing quark condensates (Cooper pairs),
is an improvement over the trivial vacuum.  The gap angle is
obtained variationally by
minimizing the vacuum ground state
$\delta \langle \Omega \arrowvert H - E \arrowvert \Omega \rangle = 0$
yielding the gap equation which is similar to the Schwinger-Dyson
equation for the quark self-energy in the rainbow approximation

\begin{equation} \label{gap3d}
k s_k - m_q c_k = \frac{2}{3} \int \frac{d\vec{p}}{(2\pi)^3}
\hat{V}(\arrowvert \vec{k}
-\vec{p}\; \arrowvert ) [s_k c_p \hat{k} \cdot \hat{p} -
s_p c_k],
\end{equation}

\noindent
where $\hat{V}(\arrowvert \vec{k}-\vec{p}\; \arrowvert ) = -8\pi\sigma/
\arrowvert \vec{k}-\vec{p}\; \arrowvert^4$ is the linear potential in
momentum space.
The solution $s_k = \sin \phi(k)$, $c_k = \cos \phi(k)$ also provides the
quark condensate $\langle \overline{q} q \rangle $
\begin{equation}
\langle \overline{q} q \rangle =
\langle \Omega \arrowvert \overline{\Psi}(0) \Psi(0) \arrowvert \Omega
\rangle = - \frac{3}{\pi^2} \int dp \;p^2\;s_p,
\end{equation}
in the BCS vacuum. From the gap equation at large $k$,
$s_k \rightarrow m_q/\sqrt{{m_q}^2 + k^2}$, yielding a quadratically
divergent condensate for non-zero current quark mass which must be
renormalized \cite{NCSU2}.  For
$m_q = 0$ we compute
$\langle
\overline{q} q \rangle \simeq -(113 \;MeV)^3$. We also
added the Coulomb interaction with a reasonable cutoff and found a
slight improvement to 119, in agreement with the more elaborate,
renormalized
result of \cite{NCSU2} but still substantially less than
lattice theory results ($\approx 250$).  The BCS vacuum also exhibits
spontaneous chiral symmetry breaking resulting in a constituent quark mass,
$\hat {m}_q$, which can not be extracted from the gap energy, $\epsilon_k$,

\begin{equation}
\epsilon_k = m_q s_k + k c_k - \frac{2}{3} \int \frac{d\vec{p}}{(2\pi)^3}
\hat{V}(\arrowvert \vec{k}
-\vec{p}\; \arrowvert ) [c_p c_k \hat{k} \cdot \hat{p} +
s_p s_k],
\end{equation}
since $\epsilon_k$ diverges for $k \rightarrow 0$.  Following previous
prescriptions \cite {Orsay,Lisbon,Adler}, we introduce the running dynamic
mass, $\hat {m}_q^{dyn}(k)$, by
$s_k = \hat{m}_q^{dyn}(k)/\sqrt{\hat{m}_q^{dyn}(k)^2
+ k^2}$ and obtain
$\hat {m}_q$ from the slope of the gap angle near zero momentum
($\hat {m}_q \approx \hat {m}_q^{dyn}(0))$.
This yields $\hat{m}_{u/d} \approx 80 $ $MeV$ for  u, d flavors
with $m_u = m_d = 5 \;MeV$ and $\hat{m}_s \approx 250 $ $MeV$ for the s
quark with $m_s = 150 $ $MeV$.  Again these values are somewhat lower than
used in phenomenological quark models indicating a more sophisticated
vacuum  is needed which is provided by the RPA as detailed below.

Using these quasi-particle creation operators we now address excited meson
states and first construct the TDA Fock space wavefunction built on the
BCS vacuum.  For a meson with quantum numbers $nJ^\pi$
(radial/node number $n$, total angular momentum $J$ and
parity $\pi$) the leading
$q\overline{q}$ state is given by

\begin{equation}
\arrowvert \Psi^{nJ^\pi} \rangle =\sum_{\lambda \mu}\int
\frac{d\vec{k}}{(2\pi)^3}
\Psi ^{nJ^\pi}_{\lambda
\mu}(\vec{k}) B^{\dagger}_{\lambda}(\vec{k})
D^{\dagger}_{\mu}(-\vec{k})
\arrowvert
\Omega \rangle,
\end{equation}

\begin{equation}
\Psi^{nJ^\pi}_{\lambda \mu}(\vec{k}) =
\sum_{LSm_Lm_S} \langle L m_L S m_S
\arrowvert J m_J \rangle
(-1)^{\frac{1}{2}+\mu} \langle \frac{1}{2} \lambda \frac{1}{2} -\mu
\arrowvert S m_S \rangle
Y^{m_L}_L(\hat{k}) \psi^{nJ^\pi}_{LS}(k).
\end{equation}
Diagonalizing $H$ in this model space generates the TDA
equation of motion (analogous to the Bethe-Salpeter equation)

\begin{equation} \label{TDA} \langle \Psi^{nJ^\pi} \arrowvert [\hat{H},
B_{\alpha}^{\dagger}D_{\beta}^{\dagger}]
\arrowvert \Omega \rangle= (E_{nJ^\pi}-E_0)\Psi^{nJ^\pi}_{\alpha \beta}.
\end{equation}

\noindent
Very significantly, the relativistic structure of our effective interaction
contains an important spin dependence. This is revealed more clearly in the
TDA partial-wave equations for a meson in state ${nJ^\pi}$ with mass
$M_{nJ^\pi}$ having quantum numbers
$L$ (orbital) and $S$ (total spin)

\begin{equation}
\label{Tammgeneral}
(M_{nJ^\pi}-2\epsilon_k)\psi^{nJ^\pi}_{LS}(k)=\int_0^\infty
\frac{dp\;p^2} {12\pi^2} \; K^{J^\pi}_{LS}(k,p)
\;\psi^{nJ^\pi}_{LS}(p),
\end{equation}
with kernel $K^{J^\pi}_{LS}(k,p)$ structure:
1) \underline{pseudoscalar} ($J^{\pi} = 0^-$), $2( c_k c_p \hat{V}_1
+(1+s_ks_p)\hat{V}_0)$;
2) \underline{scalar} $(J^\pi = 0^+)$, $2( c_k c_p \hat{V}_0
+(1+s_ks_p)\hat{V}_1)$;
3) \underline{vector} $(J^\pi = 1^-)$, $2c_kc_p\hat{V}_1
+(1+s_k)(1+s_p)\hat{V}_0
+(1-s_p)(1-s_k)(\frac{4}{3}\hat{V}_2-\frac{1}{3}\hat{V}_0)$;
4) \underline{pseudovector} $(J^\pi = 1^+)$, $c_kc_p (\hat{V}_0+\hat{V}_2)
+2(1+s_ps_k)\hat{V}_1 $.
Here $\hat{V}_i$ is the angular integral over $x = \hat{k}\cdot\hat{q}$
of $\hat{V}(\arrowvert \vec{k}-\vec{q}\; \arrowvert)$ with powers $x^i$.

The TDA spectrum is given in Fig. 1 (dotted lines) for the
pseudoscalar and vector mesons.  Note that in this model the $\rho$  and
$\omega$ states are degenerate and there is no dynamic flavor mixing
(the standard $SU_{F}(3)$ flavor mixings have been adopted for $\eta$,
$\eta'$, $\omega$ and $\phi$).  Considering that $\sigma$ is the only, but
fixed, parameter in this model, the spectrum is
qualitatively reasonable with the exception of the ground state
pion.  From model calculation the spin splitting between the $\rho$ and
$\pi$ is about $200$ $MeV$, insufficient to describe the roughly $600$ $MeV$
observed difference \cite{PDG98}.  This shortcoming is due to the
inability of the TDA to properly include constraints from chiral symmetry.

Finally, we formulate the RPA \cite{Ring} and in analogy
to nuclear physics applications generalize the meson creation operator,
$Q_{nJ^\pi}^{\dagger}$,
for state $nJ^\pi$ to both create and destroy $q\overline{q}$ pairs
with an improved vacuum satisfying $Q_{nJ^\pi}\arrowvert \Omega _{RPA }
\rangle=0
$ and containing quark correlations beyond the BCS.
To obtain the RPA equations of motion we make use of the quasi-boson
approximation in which pairs of operators, $BD$, are treated as boson
operators. For the important pseudoscalar ($J^\pi = 0^-)$ meson channel we
obtain the coupled partial-wave RPA equations

\begin{eqnarray*}
2\epsilon_k X^{n}(k) + \frac{1}{3} \int_0 ^{\infty} \frac{dp\;p^2}
{(2\pi)^2} \left[ X^{n}(p)F(k,p) + Y^{n}(p) G(k,p)
\right] = M_{n} X^{n}(k),
\\
2\epsilon_k Y^{n}(k) + \frac{1}{3} \int_0 ^{\infty} \frac{dp\;p^2}
{(2\pi)^2} \left[ Y^{n}(p)F(k,p) + X^{n}(p) G(k,p)
\right] = -M_{n} Y^{n}(k),
\end{eqnarray*}
where

\begin{eqnarray}
F(k,p)=2c_pc_k \hat{V}_1 + 2(1+s_ps_k)\hat{V}_0,\\
G(k,p)=2c_pc_k \hat{V}_1 - 2(1-s_ps_k)\hat{V}_0,
\end{eqnarray}
with similar expressions for the other meson channels.  The RPA spectrum
(dashed lines) is summarized in Fig. 1.  Notice the improvement with
only the pseudoscalar states ($\pi$ and $\eta$)
shifted downward from the TDA.  Related, we now also
obtain the correct chiral limit for the pion mass. As expected from
chiral arguments concerning the Goldstone boson nature of the pion, the RPA
pseudoscalar mass (pure u/d flavor) approaches zero for $m_q \rightarrow 0$
as illustrated in Fig. 2 (solid curve).  Appropriately, the scalar meson,
$f_0$, mass (dotted curve) converges to a non-zero value in the same limit
(the scalar and pseudovector meson spectrums are also reasonable but not
shown due to space limitations). Hence the RPA clarifies the major source
of mass splitting between the
$\pi$ and
$\rho$ -- roughly 200 $MeV$ from spin dependence, as in the TDA,
but a much larger amount, about 400 $MeV$, due to chiral symmetry
constraints.

The RPA also provides impressive improvement in the quark condensate and
pion decay constant.  Performing an expansion of the RPA vacuum in
powers of
boson operators and keeping only leading corrections
exciting up to two mesons from the BCS vacuum, we obtain
a much larger condensate. For zero current quark mass,
$\langle \overline{q}q\rangle \simeq -(300MeV)^3$, in
better agreement with the currently accepted value. Interestingly, this
overestimation of ground state correlations seems to be characteristic of
RPA, as previously documented in other fields of physics
\cite{Ring}.  Finally, in the chiral limit we compute the pion decay
constant
$f_{\pi} = 120$ $MeV$, in reasonable agreement with data ($93$ $MeV$) and
substantially better than the TDA value of $10$ $MeV$.  Related and also
noteworthy, we have numerically verified the generalized
Gell-Mann-Oakes-Renner relation

\begin{equation}
-2m_q\langle \overline{q}q\rangle= \sum_{n} M_n^2 f_n^2,
\end{equation}
by independently computing terms on both sides of the equation. The
excited pseudoscalar states negligibly contribute as their $f_n$ are
suppressed in the chiral limit.

Summarizing, we have documented the utility of several many-body techniques,
especially the RPA, for investigating the QCD structure of hadrons.  In
conjunction with our previous glueball study we have also further
established our effective Hamiltonian and many-body approach as, with
just the independent lattice parameter $\sigma$,
the semi-quantitative features of the vacuum, meson and glueball
spectrums have all been reproduced.  Because the approximations are
controllable, this framework is amenable to systematic improvement and
should be appropriate for more challenging hadronic investigations, some in
progress, such as baryons, hadron hidden flavor (higher Fock states) and
hybrids. The details of the present calculation, as well as several issues
concerning renormalization
\cite{NCSU3}
and the $\eta - \eta'$ system
 are deferred to a subsequent, major publication.
The authors wish to recognize useful discussions with J. E. Ribeiro,
A. P. Szczepaniak and the NCSU theory group.  F. L-E. acknowledges a
SURA-Jefferson Lab fellowship.  This work is partially
supported by grants DOE DE-FG02-97ER41048 and NSF INT-9807009.

\begin{figure}
\psfig{figure=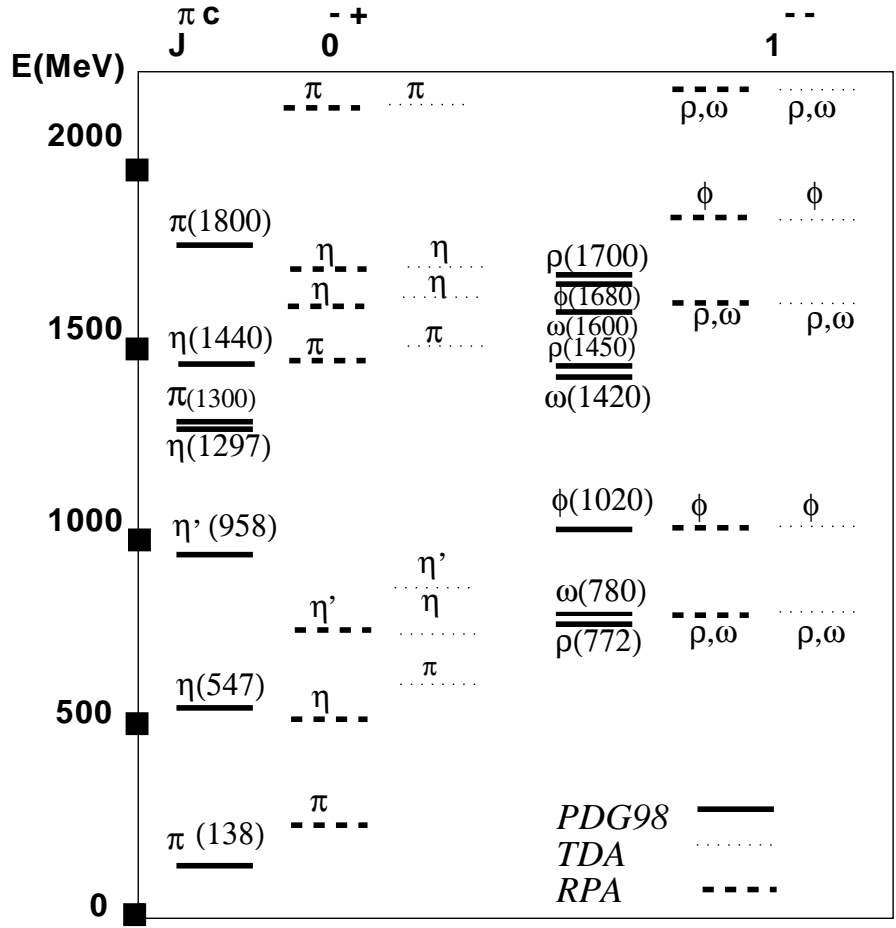,width=5in,height=6.5in}
\caption{Pseudoscalar and vector meson spectrum.  Data (solid), RPA
(dashed) and TDA(dotted).}
\label{spectrum}
\end{figure}

\newpage
\begin{figure}
\psfig{figure=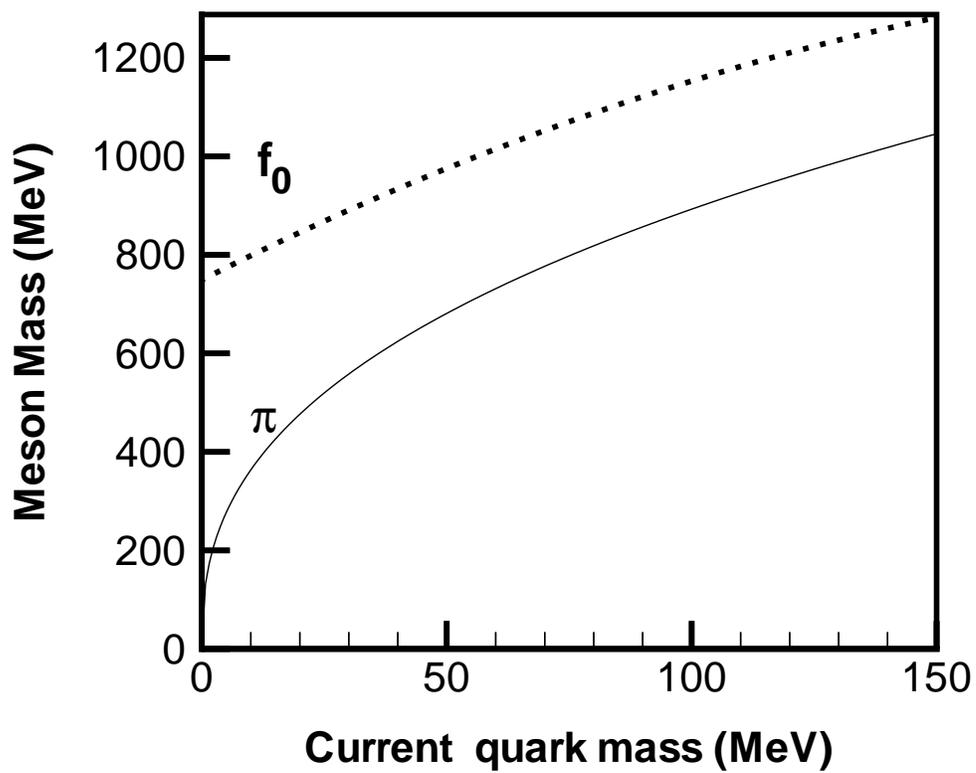,width=5in,height=7.0in}
\caption{Chiral symmetry in the RPA.
For $m_q \rightarrow 0$ the pseudoscalar (solid) but not scalar
(dotted) meson mass vanishes.}
\label{ChiralRPA}
\end{figure}

\end{document}